\documentclass[nofootinbib]{revtex4}
\usepackage{amsmath,color,fancyhdr,graphicx}
\newcommand{\GeV}{{\rm GeV}}
\newcommand{\iab}{{\rm ab}^{-1}}
\newcommand{\SM}{{\rm SM}}

\begin{document}

\title{Exploring first order phase transition in $U(1)$ extended models by complementarity between collider measurements and cosmological observations
\footnote{The talk is based on a part of Ref.~\cite{Hashino:2018zsi}}
}

\author{Toshinori Matsui}
\email{matsui@kias.re.kr}
\affiliation{School of Physics, Korea Institute for Advanced Study, \\
85 Hoegiro, Dongdaemun-gu, Seoul 02455, Republic of Korea}

\begin{abstract}
 We consider models with the $U(1)_X$ gauge symmetry, which is spontaneously broken by dark Higgs mechanism.
 We discuss patterns of the electroweak phase transition and detectability of gravitational waves (GWs) when strongly first order phase transition (1stOPT) occurs. 
 It is pointed out that the collider bounds on the properties of the discovered Higgs boson exclude a part of parameter space that could otherwise generate detectable GWs. 
 We find that GWs produced from multi-step PT can be detected at future observations such as LISA and DECIGO if the dark photon mass is $m_X^{} \gtrsim 25~\GeV$ with the $U(1)_X^{}$ gauge coupling being $g_X^{} \gtrsim 0.5$.
 In addition, we show that most of the parameter regions can be covered by precision measurements of various Higgs boson couplings and direct searches for the singlet scalar boson at future collider experiments.
 Furthermore, we expect the complementarity of the detection of GW observations from the strongly 1stOPT, collider bounds and dark photon searches in the models of the dark gauge symmetry.
\end{abstract}

\maketitle
\thispagestyle{fancy}

\section{Introduction}
\label{sec:intro}
 
 
 In models extended from the standard model (SM) Higgs sector, strongly first order phase transition (1stOPT) can be realized.
 We emphasize that the nature of electroweak phase transition (EWPT) can be probed by exploring the Higgs sector at ongoing and future experiments.
 Models predicting significant deviations in various Higgs boson couplings can be tested at the LHC~\cite{CMS:2013xfa} as well as at future lepton colliders including, the International Linear Collider (ILC)~\cite{ILC}, the Compact Linear Collider (CLIC)~\cite{CLIC} and the Future Circular Collider of electrons and positrons (FCC-ee)~\cite{FCC-ee}.

 On the cosmological side, the strongly 1stOPT that occurs in the early Universe produces stochastic gravitational waves (GWs).
  In future, planned space-based interferometers such as LISA~\cite{Seoane:2013qna}, DECIGO~\cite{Kawamura:2011zz} and BBO~\cite{Corbin:2005ny} will survey GWs in the millihertz to decihertz range, which is the typical frequency of GWs from the 1stOPT at the electroweak scale.

 Among various extensions of the Higgs sector, we here focus on a model with gauged dark $U(1)_X^{}$ symmetry including the $U(1)$ gauge kinetic mixing term.
 In general, $U(1)$ extended models are also testable at various experiments for the dark photon search.

\section{Model with dark $U(1)_X$  gauge symmetry}
\label{sec:model}

 We consider a model with a dark sector where the $U(1)_X^{}$ Abelian gauge symmetry is spontaneously broken by the so-called dark Higgs mechanism. 
 We introduce a complex scalar $S$ with $U(1)_X^{}$-charge $Q_X^{}$ and the $U(1)_X^{}$ gauge field (dark photon) $X_\mu^0$. 
 In generic, there appears the gauge kinetic mixing term between the $U(1)_X^{}$ gauge boson $X_\mu^0$ and the hypercharge $U(1)_Y^{}$ gauge boson $B_\mu^{}$~\cite{Holdom:1985ag}, and the Lagrangian is given by (e.g. Ref.~\cite{Addazi:2017gpt})
\begin{align}
{\cal L} = - \frac{1}{4} X_{\mu\nu} X^{\mu\nu} - \frac{\epsilon}{2} X_{\mu \nu} B^{\mu \nu} + |D_\mu S|^2 - V_0
\label{eq:lagrangian}
\end{align}
where $X_{\mu\nu}=\partial_\mu X_\nu^0 - \partial_\nu X_\mu^0$ and
$B_{\mu\nu}^0=\partial_\mu B_{\nu}^0 - \partial_\nu B_{\mu}^0$, \
 and the covariant derivative is defined as $D_\mu = \partial_\mu + i g_X Q_X X_\mu^0$.
Here, the Higgs potential is given by 
\begin{align}
V_0^{} =
-\mu_\Phi^2|\Phi|^2
-\mu_S^2 |S|^2
+\lambda_\Phi^{} |\Phi|^4
+\lambda_S^{} |S|^4
+\lambda_{\Phi S}^{} |\Phi|^2 |S|^2.
\label{eq:full_theory}
\end{align}
 We normalize the $U(1)_X^{}$ charge of $S$, $Q_S^{} \equiv Q_X^{}(S)$; $Q_S^{}=1$.
 Since the viable parameter range for $\epsilon$ is too small to affect PT~\cite{Addazi:2017gpt}, we focus on the rest six parameters, i.e. $\mu^2_\Phi, \mu^2_S, \lambda_\Phi^{}, \lambda_S^{}$, $\lambda_{\Phi S}^{}$ and $g_X^{}$.

 After the EW symmetry breaking, the two Higgs multiplets can be expanded as
$\Phi=(
w^+, \frac{1}{\sqrt{2}}(v_\Phi+\phi_\Phi+i z^0)
)^T, 
S=\frac{1}{\sqrt{2}}(v_S+\phi_S+i x^0)$, 
where $v_\Phi$ and $v_S$ are the corresponding vacuum expectation values (VEVs). 
 The Nambu-Goldstone modes $w^\pm$, $z^0$ and $x^0$ are absorbed by the gauge bosons $W^\pm_\mu$, $Z^0_\mu$ and $X_\mu^0$.
 The mass of $X_\mu^0$ is $m_{X} = g_X |Q_S| v_S$ (see also Ref.~\cite{Farzan:2012hh}).
 The interaction basis state ($\phi_\Phi^{}$, $\phi_S^{}$) is diagonalized to their mass eigenstate ($h$, $H$) through the rotation matrix with $c_\theta\equiv\cos\theta$ and $s_\theta\equiv\sin\theta$.
 We take $v_\Phi^{}(= 246~\GeV)$, $m_h^{}(=125~\GeV)$, $m_H^{}$, $\theta$, $g_X$ and $m_X$ as input parameters.
 The tree-level interactions of $h$ and $H$ with the SM gauge bosons $V$(=$W_\mu^\pm$, $Z_\mu^0$) and with the SM fermions  $F$ are given by
${\cal L}_{\Phi V V, \Phi FF} = (h c_\theta-H s_\theta) \{(2 m_W^2/v_\Phi) W^+_\mu W^{- \mu}+ (m_Z^2/v_\Phi)  Z^0_\mu Z^{0 \mu} -\sum_F (m_F/v_\Phi)  \bar{F} F \}$. 
 The couplings of $h$ is normalized by the corresponding SM ones are universally given by
\begin{align}
\kappa \equiv 
\frac{g_{h VV}}{g_{h VV}^\SM} =\frac{g_{h FF}}{g_{h FF}^\SM}=c_\theta. 
\end{align}

\section{Numerical results}
\label{sec:results}

\begin{figure}[t]
\centering
\includegraphics[width=0.57\textwidth]{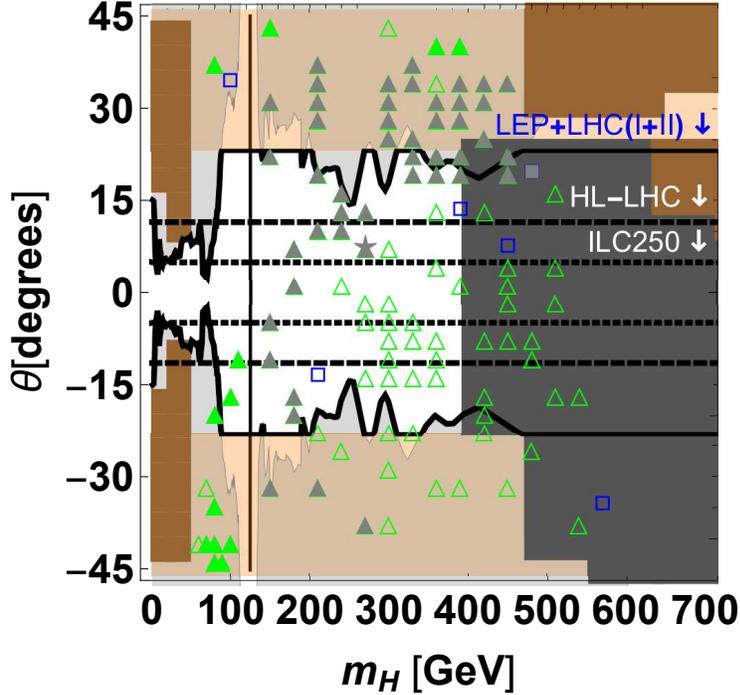}
 \caption{\label{fig:200+1_1}
 Types of multi-step PT on the $(m_H^{},\theta)$ plane for the benchmark point $m_X=200~\GeV$ and $g_X=2$.
 Parameter sets predicting one-step PT with 1st order are marked with blue closed square, one-step PT with 2nd order with blue open square, two-step PT where both transitions are 1st order with green closed star, two-step PT where the latter one is 1st order with green closed triangle, and two-step PT where the former one is 1st order with green open triangle.
 The gray plots are insensitive at future GW observations.
 The colored regions are excluded by perturbative unitarity (black), vacuum stability (brown).
 The black lines show the combined exclusion limit obtained by $\kappa_Z^{}$ measurement (orange) and direct searches for $H$ (gray).
 The black dashed lines and dotted lines show the expected accuracy of $\kappa$ measurements at HL-LHC (14~TeV, 3~$\iab$) and ILC (250~GeV, 2~$\iab$).
 }
\end{figure}

 In Fig.~\ref{fig:200+1_1}, various types of multi-step PT that predict first order EWPT are marked with colored plots.
 The gray plots are insensitive at future GW observations, LISA~\cite{Caprini:2015zlo} and DECIGO~\cite{Kawamura:2011zz}.
 We find that there are still detectable regions satisfying the current collider constraints shown below.

 The measurements of the Higgs boson decay into weak gauge bosons give constraints on the $hVV$ couplings as $\kappa_Z^{}=1.03^{+0.11}_{-0.11}$ and $\kappa_W^{}=0.91^{+0.10}_{-0.10}$ from the ATLAS and CMS combination of the LHC Run-I data (68\% CL)~\cite{TheATLASandCMSCollaborations:2015bln}.
 In our numerical analysis, we take the 68\% CL bound $\kappa_Z^{}>0.92$ as the lower bound on the mixing angle, namely $|\theta| \leq 23.1^\circ$. 
 The exclusion limits from the direct searches for the $H$ boson at the LEP and LHC Run-II are examined in Ref.~\cite{Robens:2015gla}.  
 We will show that a large portion of the model parameter space where strongly 1stOPT and detectable GW signals are possible is excluded by the collider bounds on the Higgs bosons discussed above.

 The expected accuracy of the measurements of the Higgs boson couplings are also displayed in Fig.~\ref{fig:200+1_1} as follows. 
 The high-luminosity (HL)-LHC with $\sqrt{s}=14$~TeV and $L=3~\iab$ can constrain $\Delta \kappa_V$ with an accuracy of $2\%$~\cite{CMS:2013xfa}.
 Future $e^+e^-$ colliders can considerably ameliorate the precision.
 The stage of the ILC with $\sqrt{s}=250$~GeV and $L=2~\iab$ can limit $\Delta \kappa_W^{}$ to 1.8\% and $\Delta \kappa_Z^{}$ to 0.38\%~\cite{Fujii:2017vwa}.
 In addition, the limit obtained from direct searches for the $H$-boson at future colliders are discussed in Ref.~\cite{Chang:2017ynj} for the small mass region and in Ref.~\cite{Carena:2018vpt} for the large mass region.

\begin{figure}[t]
\centering
  \includegraphics[width=0.57\textwidth]{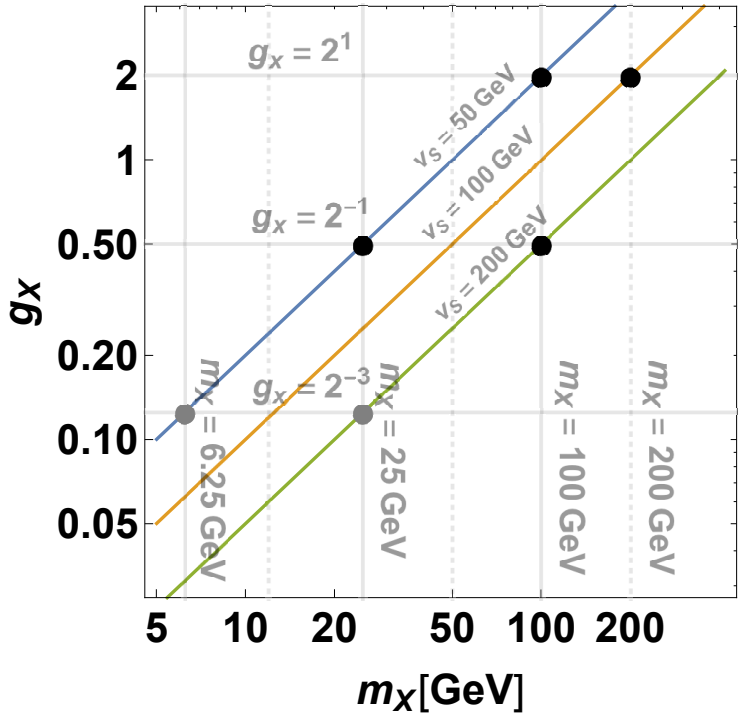}
   \caption{\label{fig:benchmark}
 The six benchmark points are blobbed on the ($m_X, g_X$)-plane.
 The solid lines show the cases of $v_S$=50~GeV (blue), 100~GeV (orange), 200~GeV (green) from the top.
 The gray plots in the small $g_X$ region shows that any parameters on the $(m_H^{},\theta)$ cannot reach to the sensitivity of future GW observations.
  }
\end{figure}

 In Fig.~\ref{fig:benchmark}, our numerical results about the EWPT and GW signals for the six benchmark points are defined.
 Scanning the parameter region with the collider bounds on the Higgs boson properties into consideration, we have found that GW signals are detectable only for larger dark photon mass, say $m_X^{}\gtrsim {\cal O} (25-100)~\GeV$.
 As shown in Ref.~\cite{He:2017zzr}, the recent data from LHCb~\cite{Aaij:2017rft} and LHC Run-II~\cite{Aaboud:2017buh} give constraints on $\epsilon$, which is roughly smaller than $10^{-2}$ at least, for the mass regions $10.6~\GeV <m_X <70~\GeV$ and $150~\GeV <m_X <350~\GeV$, respectively.
 It is also shown that the mass region $20~\GeV <m_X <330~\GeV$ can be constrained at future lepton colliders in Ref.~\cite{He:2017zzr}.
 We expect that 1stOPT with such a heavy dark photon will be tested by synergy between future GW observations and dark photon searches.

\section{Conclusions}
\label{sec:conclusions}

 We have comprehensively explored models with a dark photon whose mass stems from spontaneous $U(1)_X^{}$ gauge symmetry breaking by the nonzero VEV of the dark Higgs boson $S$ in light of the patterns of PT and the detectability of GWs from strongly 1stOPT as well as various collider and theoretical bounds.
 After imposing these constraints on the model parameter space, we have found that GWs produced from multi-step PT can be detected at future observations such as LISA and DECIGO, if the dark photon mass is $m_X^{} \gtrsim 25~\GeV$ with the $U(1)_X^{}$ gauge coupling being $g_X^{} \gtrsim 0.5$.
 We have found that the parameter regions predicting detectable GWs are covered by the measurement of $\kappa^{}$ at future colliders including the HL-LHC and ILC.

\acknowledgments

 This work is based on the collaboration with Katsuya Hashino, Mitsuru Kakizaki, Shinya Kanemura and Pyungwon Ko. I would like to thank them for their support.

\end{document}